\documentclass[epj]{svjour}
\usepackage{graphicx} % Graphics
\usepackage{booktabs} % Table Commands
\usepackage{hyperref} % Hyperlinks
\usepackage{times} % PostScript fonts
\begin{document}
\title{Sensitivity of the r-process to Nuclear Masses}
\author{S. Brett\inst{1,2} \and I. Bentley\inst{1} \and N. Paul\inst{1} \and R. Surman\inst{3} \and A. Aprahamian\inst{1}}

\institute{Dept. of Physics, University of Notre Dame, Notre Dame, IN 46556 \and Dept. of Physics, University of Surrey, Guildford, Surrey, GU2 7XH, United Kingdom \and Dept. of Physics, Union College, Schenectady, NY 12308}

\date{Received: date / Revised version: date}
\abstract{
The rapid neutron capture process (r-process) is thought to be responsible for the creation of more than half of all 
elements beyond iron. The scientific challenges to understanding the origin of the heavy elements beyond iron lie in both 
the uncertainties associated with astrophysical conditions that are needed to allow an r-process to occur and a vast lack 
of knowledge about the properties of nuclei far from stability. There is great global competition to access and measure 
the most exotic nuclei that existing facilities can reach, while simultaneously building new, more powerful accelerators 
to make even more exotic nuclei. This work is an attempt to determine the most crucial nuclear masses to measure 
using an r-process simulation code and several mass models (FRDM, Duflo-Zuker, and HFB-21). The most important 
nuclear masses to measure are determined by the changes in the resulting r-process abundances.  Nuclei around the 
closed shells near N=50, 82, and 126 have the largest impact on r-process abundances irrespective of the mass models used.
 % end of PACS codes
} %end of abstract

\maketitle
Basic properties of nuclei, such as their binding energies per nucleon allow the synthesis of the elements up to 
approximately iron via fusion reactions in stars from the lightest elements created by the Big Bang. However, the 
abundances of elements in our solar system contain a substantial number of nuclei well beyond 
iron~\cite{Arla99,GA10,Lo03}. The origins of these nuclei are entangled in complexity since the heavier elements are 
thought to be made via both slow- and rapid- neutron-capture processes (s- and r-processes) \cite{Sneden2008}.  The 
s-process leads to a network of nuclei near stability while the r-process allows the production of nuclei with 
increasing neutron numbers much further from stability, producing neutron-rich nuclei. The astrophysical scenarios in 
which the s-process can take place have been identified, but a potential site for the r-process is still 
unresolved\cite{Arnould2007}. The challenge for astrophysical science today is to understand the conditions that 
would provide a major abundance of neutrons and lead to successive captures before the nucleus has a chance to decay; 
while on the nuclear side, the challenge is to determine the physics of nuclei far from stability where the range and 
impact of the nuclear force is less well known \cite{KG99,Arnould2007}.

There have been a number of astrophysical scenarios suggested as possible sites for the r-process. Some of the most 
promising sites include the neutrino driven wind from core-collapse supernovae \cite{WW94}, two-neutron star-mergers 
\cite{Freiburghaus1999}, gamma-ray bursts \cite{Surman2006}, black-hole neutron star mergers \cite{Surman2008}, 
relativistic jets associated with failed supernovae \cite{Fujimoto2006} or magnetohydrodynamic jets 
from supernovae \cite{Nishimura2006}.

The r-process proceeds via a sequence of neutron captures, photodissociations and $\beta$ decays.  Simulations
of the r-process therefore require tabulations of $\beta$-decay lifetimes, neutron capture rates and
neutron separation energies; photodissociation rates are determined from the capture rates and separation 
energies by detailed balance \cite{FCZ67}:
\begin{equation} 
\lambda_\gamma(Z,A) \propto T^{3/2} \exp\left[-{\frac{S_n(Z,A)}{kT}}\right] \langle \sigma v \rangle_{(Z,A-1)}
\label{photo}
\end{equation} 
In the above expression, $T$ is the temperature, $\langle \sigma v \rangle_{(Z,A-1)}$ is the thermally-averaged value of
the neutron capture cross section for the neighboring nucleus with one less neutron, and $S_{n}(Z,A)$ is the neutron
separation energy---the difference in binding between the nuclei $(Z,A)$ and $(Z,A-1)$.  Nuclear masses are crucial inputs
in theoretical calculations of each of these sets of nuclear data.

One way to assess the role of nuclear masses in the r-pro\-cess is to choose two or more mass models, calculate 
all of the relevant nuclear data with the mass model consistently, and then run r-process simulations with the different 
sets of global data.  Such comparisons are quite valuable and examples include Ref.~\cite{Wan04,Far10,Arc11}.  
Our approach here is quite different.  We instead focus on the sensitivity of the r-process to the \emph{individual} 
neutron separation energies within a given mass model, as they appear in Eqn.~\ref{photo}, in an attempt to determine the nuclei that have the 
greatest impact on the overall r-process abundances and, in turn, identify the most crucial measurements to be made. This 
is the first time that such an attempt has been made and the results could potentially be of great significance to both 
nuclear and astrophysical science.

The study of radioactive nuclei far from stability approaching the r-process path is one of the global research frontiers 
for nuclear science today. New facilities are being developed in the USA (CARIBU at ANL, NSCL and FRIB at MSU), in Europe 
(ISOLDE at CERN), in France (SPIRAL II at GANIL), in Finland (Jyvaskyla), in Germany (FAIR at GSI Darmstadt), in Japan 
(RIKEN), in China (BRIF,CARIF in CIAE Beijing), and in Canada (ISAC at TRIUMF). The overarching question for this global 
effort in nuclear science is which measurements need to be made \cite{Sch08}.

This study used a fully dynamical r-process nuclear network code \cite{Wa94}. Inputs to the simulation code include a seed 
nucleus, neutron density, temperature and dynamical timescale descriptive of a given astrophysical scenario. In 
addition, $\beta$ decay rates, neutron capture rates and neutron separation energies are the inputs for the nuclear 
properties.  The simulation processes neutron captures, photodissociations, $\beta$-decays, and $\beta$-delayed 
neutron emissions from the start of the r-process through freezeout and the subsequent decay toward stability 
\cite{Me02}.  Fission, while important in some astrophysical scenarios, is not significant for the the conditions 
used here and so is not included.

\begin{figure}
\includegraphics[width=8.5cm]{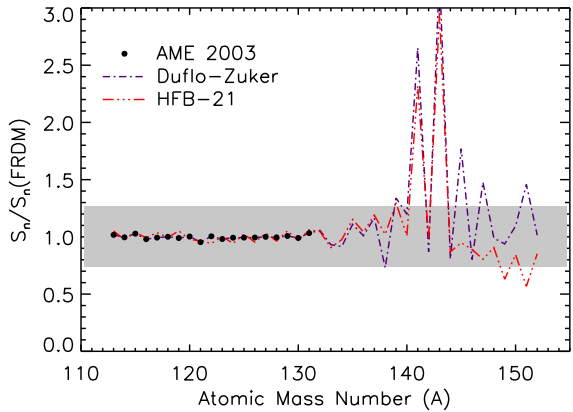}
\caption{Comparison of the separation energies from Duflo-Zuker \cite{DZ95}, HFB-21 \cite{Gor10}, and the experimental
masses from \cite{Au03} to the FRDM \cite{MN95} values for the tin isotopes.}
\label{fig:MM}
\end{figure}

All the calculations are done for the same initial astrophysical conditions.  The astrophysical scenario used in our 
simulations was based on the H or high frequency r-process suggested by Qian et al. \cite{Qi98}, with an initial 
temperature of $T_{9} = 1.5$ and an initial density of $3.4\times 10^{2}$ g/cm$^{3}$.  We take the temperature and 
density to decline exponentially as in \cite{QiW96} with a dynamical timescale of 0.86 s. While Qian specifies a seed 
of $^{90}$Se and a neutron to seed ratio ($N_n/N_{seed}$) of 86 \cite{Qi98}, here a lighter seed of $^{70}$Fe is 
chosen, which results in $N_n/N_{seed}=67$ when the electron fraction is kept consistent with Qian ($Y_{e}=0.190$).

The nuclear data inputs include beta decay rates from \cite{MPK03} and neutron capture rates from \cite{RT00}, both 
calculated with Finite Range Droplet Model (FRDM) masses.  The measured values of $S_n$ come from the Audi Mass 
Evaluation 2003 \cite{Au03}. For the remaining nuclei, we used the $S_n$ values resulting from the calculated mass 
values in the FRDM \cite{MN95}.  We subsequently varied these theoretical $S_n$ for one nucleus at a time by $\pm25\%$. 
In each case, the resulting r-process abundance curves were generated and compared against the baseline abundances 
resulting from the unchanged $S_n$ value.

The $25\%$ variation of separation energies was chosen somewhat arbitrarily. A comparison of the ratio of separation 
energies extracted from measured masses or theoretically calculated separation energies with the FRDM calculated values is 
shown for the Sn isotopes in Figure \ref{fig:MM}. This indicates that the $25\%$ value is a reasonable variation estimate 
far from stability. 

\begin{figure}
\includegraphics[width=8.5cm]{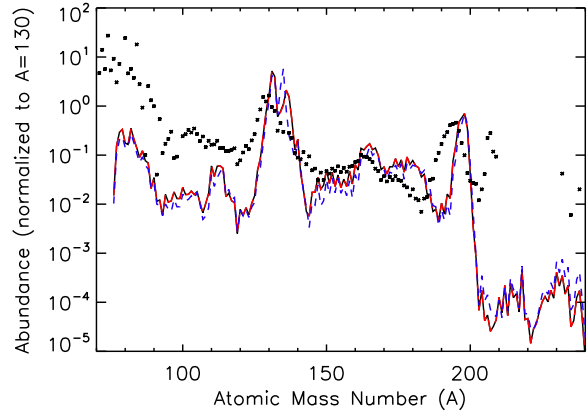}
\caption{Final r-process abundances for the baseline H-scenario \cite{Qi98} with $^{70}$Fe seed (black line) compared 
to simulations in which the neutron separation energy of $^{138}$Sn is increased (red long-dashed line) or decreased (blue 
short-dashed line) by $25\%$.  The calculated abundances are normalized to the solar r-process abundances of Sneden et al. 
\cite{Sneden2008} (points) at $A=130$.}
\label{fig:138SnwSolar}
\end{figure}

An example of the resulting abundance patterns is shown in Figure \ref{fig:138SnwSolar}, where the baseline pattern is compared to the
final abundance patterns produced by simulations in which the separation energy of $^{138}$Sn was increased or decreased by $25\%$. 
This comparison can be quantified by summing the differences in the final mass fractions:
\begin{equation}
\label{eqn:1}
F_{\pm}=100\sum_{A} \vert X_{baseline}(A)- X_{\pm\Delta S_n}(A)) \vert, 
\end{equation}
where $X(A)=AY(A)$ is the mass fraction of nuclei with mass number $A$ (such that $\sum_{A} X(A)=1$), and the sum of $A$ 
ranges over the entire abundance curve.  This quantity is largest when the curves differ near the peak abundances, 
giving preference to those regions.

\begin{figure}
\includegraphics[width=8.7cm]{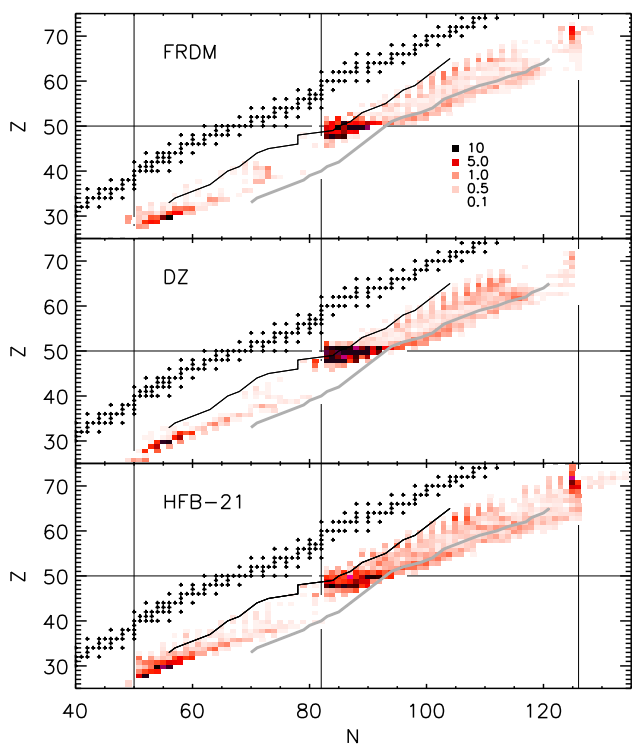}
\caption {Comparison of the sensitivity to mass values determined by Equation 2. The separation energies far from stability were
generated by the FRDM \cite{MN95}, Duflo-Zuker \cite{DZ95}, and HFB-21 \cite{Gor10}. The scale is from white to dark red, indicating
regions with a small change to a substantial change in the resulting abundances.  For reference, stable nuclei have been
included as black crosses and the magic numbers have been indicated by thin lines. Superimposed on the sensitivity results are the limits of accessibility
by CARIBU \cite{SavPar05} and the proposed FRIB intensities \cite{TarHau12}. In both cases, we have plotted the conservative limits of what can be produced 
and measured in mass measurements.} 
\label{fig:sens}
\end{figure}

The values of $F=(F_{+}+F_{-})/2$ are calculated for 3010 nuclei from $^{58}$Fe to $^{294}$Fm. Figure 
\ref{fig:sens} shows the nuclei whose separation energy variations result in the greatest changes in the resulting 
r-process abundances. Nuclei that have the greatest impact on the r-process are those neutron rich nuclei near the closed 
shells at $Z=28$ and 50, and $N=50$, 82, and 126.

A natural question to ask is the dependence of these results on the mass model used. Therefore, similar calculations were
performed using four additional mass models, the Duflo-Zuker (DZ) \cite{DZ95}, the Extended Thomas Fermi plus Strutinsky
Integral with shell Quenching (ETFSIQ) \cite{PN96}, the Hartree-Fock-Bogoliubov (HFB-21) \cite{Gor10}, and the F-spin
\cite{Ap11} model in addition to the FRDM.  All models take advantage of very different physics ingredients to
calculate the masses of nuclei far from stability. Each of the calculations performed started with the same initial
astrophysical conditions and again varying individual separation energies by $\pm25\%$. The results are astounding.  In
each case, the nuclei with the greatest impact were generally the ones near the major closed shells independent of the
chosen mass models. Figure \ref{fig:sens} shows the resulting sensitivity plots from three of the mass models; the FRDM, DZ, and HFB-21 models. 
Nuclei near the closed shells of N=50, 82, and 126 rise above all the others in impact. The nuclei with the most impact
on the r-process abundances cluster around $^{132}$Cd and $^{138}$Sn. In this region, the nuclei are $^{131-134}$Cd,
$^{132-137}$In, $^{135-140}$Sn, $^{139,141}$Sb. There are also specific low mass nuclei such as $^{82}$Cu, $^{85}$Zn, and
$^{88}$Zn that are important.

\begin{figure}
\includegraphics[width=8.8cm]{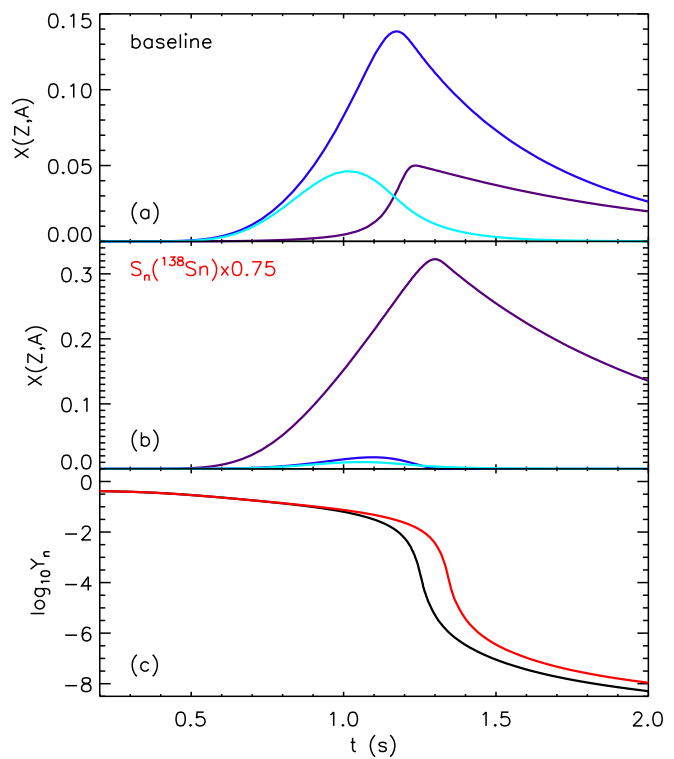}
\caption{Shows the mass fractions of $^{136}$Sn (purple), $^{138}$Sn (blue), and $^{140}$Sn (aqua) for the baseline $r$-process simulation (top panel) and the simulation with the separation energy of $^{138}$Sn decreased
by 25\% (middle panel). The bottom panel compares the neutron abundance for the two simulations (black and red lines,
respectively).}
\label{fig:mech}
\end{figure}

In trying to understand these results, we know that there are two ways that an individual neutron separation energy can 
influence the $r$-process abundance distribution. The first is a long-recognized \cite{BB57} equilibrium effect, and the 
second is an early-freezeout photodissociation effect, recently pointed out in \cite{Sur2009}. In the classic view, the 
$r$-process takes place in conditions of $(n,\gamma)$-$(\gamma,n)$ equilibrium, where abundances along an isotopic chain 
are determined by a Saha equation:
\begin{eqnarray}\nonumber
I_{00}&=& \frac{Y(Z,A+1)}{Y(Z,A)}=\frac{G(Z,A+1)}{2G(Z,A)}\left(\frac{2\pi\hbar^{2}N_{A}}{m_{n}kT}\right)^{3/2}N_{n}\\
&& \times\exp\left[\frac{S_{n}(Z,A+1)}{kT}\right]
\label{saha}
\end{eqnarray}
where the $G$s are the partition functions, $N_{n}$ is the neutron number density, and $m_{n}$ is the nucleon mass.  The relative
abundances of the different isotopic chains are then determined by the $\beta$-decay lifetimes of the most populated nuclei along each
chain.  As described in Eqn.~\ref{saha}, any change to an individual separation energy will cause a shift in the abundances along the
isotopic chain.  This can have a global impact on the final abundance pattern, particularly if the affected nucleus is highly populated
and material is shifted to a nucleus with a significantly faster or slower $\beta$-decay lifetime.  For example, consider the case of
$^{138}$Sn, a nucleus just above the $N=82$ closed shell region. In the baseline simulation,  $^{138}$Sn is the most
abundant tin isotope, and $^{136}$Sn $^{140}$Sn are much less abundant.  Their
mass fractions are shown as a function of time in Fig.~\ref{fig:mech}(a); their relative values follow those predicted by
Eqn.~\ref{saha} until about $t\sim 1.2$ s, when equilibrium begins to fail and the nuclei primarily $\beta$-decay to stability. If the
simulation is repeated with neutron separation energy of $^{138}$Sn reduced by 25\%, we see that the equilibrium abundance of this
nucleus is drastically reduced, as expected from Eqn.~\ref{saha} and shown in Fig.~\ref{fig:mech}(b).  Material is instead shifted to
$^{136}$Sn, which has a $\beta$-decay lifetime approximately 1.6 times that of $^{138}$Sn (and 5.3 times the lifetime of $^{140}$Sn,
which is also depleted by the shift). As a result, more material is stuck in the tin isotopic chain compared to the baseline
simulation, and the overall rate at which neutrons are consumed is slowed, as shown in Fig.~\ref{fig:mech}(c).  This impacts the
availability of neutrons for the whole abundance pattern and results in changes throughout the pattern.  The second
mechanism, in contrast, operates once $(n,\gamma)$-$(\gamma,n)$ equilibrium begins to fail, and individual neutron capture and
photodissociation rates become important.  Since the neutron separation energy appears in the exponential in Eqn.~\ref{photo},
photodissociation rates are quite sensitive to this quantity.  Changes in individual photodissociation rates during freezeout can
produce local shifts in abundances, which can translate into global abundance changes if they alter the late-time availability of free
neutrons.  This mechanism is described carefully in \cite{Sur2009}. Odd-$N$ nuclei, which tend to be in equilibrium only briefly if at
all, are particularly susceptible to these non-equilibrium effects.

In conclusion, this study of 3010 nuclei via an r-process simulation tested the sensitivity of the r-process abundance 
yie\-lds to the theoretical mass values of neutron rich nuclei present\-ly unknown in the laboratory from several 
different mass models, the results are shown here for three of them (FRDM\cite{MN95}, Duflo-Zuker\cite{DZ95}, and 
HFB-21\cite{Gor10}). The results are uniform and conclusive in highlighting the importance of nuclei near closed 
shells. Essentially the same set of nuclei emerge as having the highest impact on the r-process irrespective of the 
varying physics ingredients of the different mass models.  The nuclei with greatest impact on the r-process---neutron 
rich isotopes of cadmium, indium, tin, and antimony in the $N=82$ region, nickel, copper, zinc, and gallium in the 
$N=50$ region, and thulium, ytterbium, lutetium, and hafnium in the $N=126$ region---should be of highest priority to 
measure in the various exotic beam facilities around the world. Table \ref{tbl:Tab1} shows the top 25 nuclei with the 
greatest impact on the r-process for the three models.
 Since the particular isotopes of these elements that have the greatest impact can shift 
depending on the astrophysical conditions, a future paper will explore the effects of various astrophysical scenarios 
in determining the most important nuclei to measure.

\begin{table}
\begin{center}
\caption{MOST IMPORTANT NEUTRON SEPARATION ENERGIES FOR H-SCENARIO WITH $^{70}$Fe SEED \label{tbl:Tab1}}
\begin{tabular}{cccccc}
\toprule & FRDM & & DZ & & HFB-21 \\
{$^A$X} & $F$ & {$^A$X} & $F$ & {$^A$X} & $F$ \\
$^{ 138}$ Sn & 24.59 & $^{ 132}$ Cd & 36.54 & $^{ 140}$ Sn & 17.59 \\
$^{ 132}$ Cd & 22.37 & $^{ 138}$ Sn & 26.74 & $^{ 134}$ Cd & 15.77 \\
$^{ 139}$ Sn & 19.64 & $^{ 134}$ Cd & 25.96 & $^{  80}$ Ni & 12.09 \\
$^{ 137}$ Sn & 18.06 & $^{ 137}$ Sn & 23.23 & $^{  86}$ Zn & 11.85 \\
$^{ 137}$ Sb & 13.69 & $^{ 140}$ Sn & 21.79 & $^{  85}$ Zn & 11.05 \\
$^{ 140}$ Sn & 11.12 & $^{  86}$ Zn & 21.15 & $^{ 197}$ Hf & 10.62 \\
$^{  86}$ Zn & 10.24 & $^{ 139}$ Sn & 17.25 & $^{ 137}$ Sn & 10.33 \\
$^{ 135}$ Sn &  9.40 & $^{ 136}$ Sn & 16.61 & $^{ 132}$ Cd &  9.47 \\
$^{ 134}$ Cd &  8.27 & $^{ 133}$ Cd & 14.33 & $^{  84}$ Zn &  9.23 \\
$^{ 133}$ Cd &  7.72 & $^{ 135}$ Sb & 13.80 & $^{ 141}$ Sn &  8.89 \\
$^{ 131}$ Cd &  7.25 & $^{ 131}$ Cd & 13.16 & $^{ 142}$ Sn &  8.35 \\
$^{  85}$ Zn &  7.08 & $^{ 141}$ Sb & 12.25 & $^{ 136}$ Cd &  7.98 \\
$^{ 135}$ In &  6.66 & $^{ 133}$ In & 12.04 & $^{ 135}$ Cd &  7.76 \\
$^{ 141}$ Sb &  6.24 & $^{  85}$ Zn & 11.92 & $^{ 131}$ Cd &  7.63 \\
$^{ 136}$ Sn &  6.23 & $^{ 135}$ Sn & 11.54 & $^{ 196}$ Lu &  7.17 \\
$^{ 132}$ In &  5.92 & $^{ 133}$ Sn & 11.52 & $^{ 133}$ Cd &  7.12 \\
$^{ 133}$ Sn &  5.46 & $^{ 139}$ Sb & 10.77 & $^{ 137}$ In &  6.66 \\
$^{ 137}$ In &  4.77 & $^{ 135}$ In & 10.72 & $^{ 139}$ Sn &  6.00 \\
$^{ 133}$ In &  4.68 & $^{ 137}$ Sb &  9.72 & $^{ 195}$ Yb &  5.50 \\
$^{ 142}$ Sb &  4.44 & $^{ 136}$ Sb &  9.56 & $^{ 138}$ In &  5.43 \\
$^{ 197}$ Hf &  4.38 & $^{ 143}$ Sb &  9.28 & $^{ 139}$ In &  5.32 \\
$^{  89}$ Ga &  4.33 & $^{ 138}$ Sb &  8.72 & $^{  79}$ Ni &  5.23 \\
$^{ 134}$ In &  4.16 & $^{ 137}$ In &  8.14 & $^{  87}$ Ga &  5.16 \\
$^{ 139}$ Sb &  4.15 & $^{ 134}$ Sb &  7.61 & $^{ 196}$ Yb &  5.03 \\
$^{ 135}$ Sb &  4.14 & $^{ 134}$ Sn &  7.50 & $^{ 132}$ In &  5.03 \\
\toprule
\end{tabular}
\end{center}
\end{table}

This work was supported by the National Science Foundation through grant number PHY0758100 and the Joint Institute for 
Nuclear Astrophysics grant number PHY0822648.


\begin{thebibliography}{99}

\bibitem{Arla99} C. Arlandini, F. Kappeler, K. Wisshak, Astrophys. J {\bf 525}, 886 (1999).
\bibitem{GA10} N. Grevesse, M. Asplund, A. Sauval, P. Scott, Astrophysics and Space Science {\bf 328}, 179 (2010).
\bibitem{Lo03} K. Lodders, Astrophys. J {\bf 591}, 1220 (2003).
\bibitem{Sneden2008} C. Sneden, J.J. Cowan, R. Gallino, Ann. Rev. Astron. Astrophys. {\bf 46}, 241 (2008).
\bibitem{Arnould2007} M. Arnould, S. Goriely, K. Takahashi, Phys. Rep. {\bf 450}, 97 (2007).
\bibitem{KG99} K.-L. Kratz, J. G\"{o}rres, B. Pfeiffer, M. Wiescher. Journal of Radioanalytical and Nuclear Chemistry {\bf 243}, 133 (2000).
\bibitem{WW94} S.E. Woosley, J.R. Wilson, G.J. Mathews, R.D. Hoffman, B.S. Meyer, Astrophys. J {\bf 433}, 229 (1994).
\bibitem{Freiburghaus1999} C. Freiburghaus, S. Rosswog, F.-K. Thielemann, Astrophys. J {\bf 525}, L121 (1999).
\bibitem{Surman2006} R. Surman, G.C. McLaughlin, W.R. Hix, Astrophys. J {\bf 643}, 1057 (2006).
\bibitem{Surman2008} R. Surman, S. Kane, J. Beun, G.C. McLaughlin, W.R. Hix, J. Phys. G {\bf 35}, 014059 (2008).
\bibitem{Fujimoto2006} S.-I. Fujimoto, K. Kotake, S. Yamada, M.-A. Hashimoto, K. Sato, Astrophys. J {\bf 644}, 1040 (2006).
\bibitem{Nishimura2006} S. Nishimura, K. Kotake, M.-A. Hashimoto, S. Yamada, N. Nishimura, S. Fujimoto, K. Sato, Astrophys. J {\bf 642}, 410 (2006).
\bibitem{FCZ67} W. Fowler, G. Caughlan, B. Zimmerman, Ann. Rev. Astron. Astrophys. {\bf 5}, 525 (1967).
\bibitem{Wan04} S. Wanajo, S. Goriely, M. Samyn, N. Itoh, Astrophys. J {\bf 606}, 1057 (2004).
\bibitem{Far10} K. Farouqi, K.-L. Kratz, B. Pfeiffer, T. Rauscher, F.-K. Thielemann, J.W. Truran, Astrophys. J {\bf 712}, 1359 (2010).
\bibitem{Arc11} A. Arcones, G. Martinez-Pinedo, Phys. Rev. C {\bf 83}, 045809 (2011).
\bibitem{Sch08} H. Schatz, Physics Today {\bf 61}, 40 (2008).
\bibitem{Wa94} J. Walsh, Ngam.f Fortran code, Clemson University (1994).
\bibitem{Me02} B.S. Meyer, Phys. Rev. Lett. {\bf 89}, 231101 (2002).
\bibitem{DZ95} J. Duflo, A.P. Zuker, Phys. Rev. C {\bf 52}, R23 (1995).
\bibitem{Gor10} S. Goriely, N. Chamel, J.M. Pearson, Phys. Rev. C {\bf 82}, 035804 (2010).
\bibitem{Au03} G. Audi, A. H. Wapstra, C. Thibault, Nucl. Phys. A {\bf 729}, 337 (2002).
\bibitem{MN95} P. M\"{o}ller, J.R. Nix, W.D. Myers, W.J. Swiatecki, Atomic Data and Nuclear Data Tables {\bf 59}, 185 (1995).
\bibitem{Qi98} Y.-Z. Qian, P. Vogel, G.J. Wasserburg, Astrophys. J {\bf 494}, 285 (1998).
\bibitem{QiW96} Y.-Z. Qian, S.E. Woosley, Astrophys. J {\bf 471}, 331 (1996).
\bibitem{MPK03} P. M\"{o}ller, B. Pfeiffer, K.-L. Kratz, Phys. Rev. C {\bf 67}, 055802 (2003).
\bibitem{RT00} T. Rauscher, F.-K. Thielemann, Atomic Data and Nuclear Data Tables {\bf 75}, 1 (2000).
\bibitem{SavPar05} G. Savard, R. Pardo, Proposal for the $^{252}$Cf source upgrade to the ATLAS facility, Technical report, ANL (2005).
\bibitem{TarHau12} O.B. Tarasov, M. Hausmann, LISE++ development: Abrasion-Fission, Technical report, NSCL, MSUCL1300 (2005).
\bibitem{PN96} J.M. Pearson, R.C. Nayak, S. Goriely, Phys. Lett. B {\bf 387}, 455 (1996).
\bibitem{Ap11} A. Teymurazyan, A. Aprahamian, I.Bentley, N.Paul, in preparation (2012).
\bibitem{BB57} E.M. Burbidge, G.R. Burbidge, W.A. Fowler, F. Hoyle, Rev. Mod. Phys. {\bf 29}, 547 (1957).
\bibitem{Sur2009} R. Surman, J. Beun, G.C. McLaughlin, W.R. Hix, Phys. Rev. C {\bf 79}, 045809 (2009).

\end{thebibliography}
\end{document}